\documentclass[12pt,preprint]{aastex}
\begin{document}

\title{Spectroscopic Observations of the Mass Donor Star in SS~433\altaffilmark{1}}

\altaffiltext{1}{Based on observations obtained at the Gemini Observatory, which is operated by the
Association of Universities for Research in Astronomy, Inc., under a cooperative agreement
with the NSF on behalf of the Gemini partnership: the National Science Foundation (United
States), the Science and Technology Facilities Council (United Kingdom), the
National Research Council (Canada), CONICYT (Chile), the Australian Research Council
(Australia), CNPq (Brazil) and SECYT (Argentina).}

\author{T. C. Hillwig\altaffilmark{2}}
\altaffiltext{2}{Department of Physics and Astronomy, Valparaiso University, Valparaiso, IN 46383; 
todd.hillwig@valpo.edu}
\author{D. R. Gies\altaffilmark{3}}
\altaffiltext{3}{Center for High Angular Resolution Astronomy, Department of Physics and Astronomy, 
Georgia State University, P. O. Box 4106, Atlanta, GA 30302-4106; gies@chara.gsu.edu}

\begin{abstract}
The microquasar SS~433 is an interacting massive binary consisting of 
an evolved mass donor and a compact companion that ejects relativistic jets. 
The mass donor was previously identified through spectroscopic observations
of absorption lines in the blue part of the spectrum that showed Doppler 
shifts associated with orbital motion and strength variations related to
the orbital modulation of the star-to-disk flux ratio and to disk obscuration.  
However, subsequent observations revealed other absorption features that 
lacked these properties and that were probably formed in the disk gas outflow. 
We present here follow-up observations of SS~433 at orbital and precession 
phases identical to those from several previous studies with the goals 
of confirming the detection of the mass donor spectrum and providing more 
reliable masses for the two system components.  We show that the absorption 
features present as well as those previously observed almost certainly belong 
to the mass donor star, and we find revised masses of $12.3\pm 3.3$ 
and $4.3\pm 0.8 M_\odot$ for the mass donor and compact object, respectively.
\end{abstract}
\keywords{stars: individual (SS~433, V1343 Aquilae) --- X-rays: binaries
--- star: winds, outflows --- stars: individual (HD~9233) --- supergiants}

\section{INTRODUCTION}

The unique system SS~433 is an X-ray binary star that falls in the class of microquasars \citep{fab04}.  
The system exhibits relativistic jets originating from a compact object surrounded by an accretion disk.
The companion star in the system contributes only a small fraction of the total light and is thus
difficult to detect.  Because the accretion disk and its wind are the dominant source of light in the system, 
reasonable measurements have been made of the radial velocity amplitude of the compact object
\citep*[e.g.][]{fab90,gie02a}.  Models of the precessing jets and eclipses have also provided 
a very accurate system inclination \citep[78\fdg8,][]{mar89}.  Therefore, identifying the 
mass donor and accurately determining its radial velocity amplitude will lead to direct 
kinematical masses for both components.

The first potential observations of the mass donor star were reported by \citet*{gie02b} 
with additional observations published later by \citet{che05}.  These studies detected
faint absorption lines present in blue spectra of SS~433 that exhibited the 
Doppler shifts expected for the donor star.  Finding these lines is difficult, 
not only due to the small relative flux contribution of the A-star, but also to the 
active emission spectrum from the accretion disk, jets, and strong disk wind in the system.  
The most convincing observations to date of the mass donor spectrum were obtained by 
\citet{hil04}.  Their spectra suggested that the donor star has a spectral classification
of A3--7~I, and their radial velocity measurements led to component masses of 
$M_O=10.9\pm3.1~M_\odot$ and $M_X=2.9\pm0.7~M_\odot$ (where the subscripts represent 
the optical companion star {\it O} and the X-ray emitting compact object {\it X}).

This identification of the mass donor star was challenged by \citet{cha04}, \citet{bar06}, 
and \citet*{cla07}.  These authors present spectra of SS~433 that show features found 
in an A-supergiant, but that do not follow a coherent orbital radial velocity curve. 
Their observations show a larger scatter, were performed over a greater variety 
of orbital and precessional phases, and have a much lower average velocity than
those presented by \citet{hil04}.  \citet{bar06} and \citet{cla07} suggest that
these A-supergiant absorption lines arise in an accretion-driven outflow. 
Thus, their work demonstrates that not all the absorption features in the 
spectrum of SS~433 form in the photosphere of the mass donor star. 

Some of the differences between these investigations are due to the timing 
of the observations and the inherent time variability of the spectrum. 
Given the relative faintness of the mass donor star compared to the 
super-Eddington disk in the system, \citet{hil04} planned observations 
for orbital phases when the disk is partially eclipsed and 
for precessional phases when the donor star appears above the disk
plane at orbital inferior conjunction.  This strategy maximizes the 
opportunity to reduce the glare of the disk and avoid obscuration by 
the extensive disk gas.  On the other hand, \citet{bar06} made observations
at a variety of other orbital and precessional phases and made plausibility 
arguments that the donor star spectrum might appear at some of these (but
was not in fact detected).

Given the need to resolve the differing interpretations in the papers of 
\citet{hil04} and \citet{bar06}, we undertook follow-up observations 
to repeat the observations of \citet{hil04}.  If the absorption features identified 
do originate in a disk outflow, then the highly variable nature of the outflow in 
SS~433 suggests that repeat observations will show significant scatter from prior 
observations, if in fact the absorption features are still present.  
However, if the absorption lines originate from the mass donor star, then the highly stable 
and periodic nature of the orbit should produce observations consistent with the data 
of \citet{hil04}.

\section{Observations and Reductions}

Observations were acquired with GMOS on the Gemini North Telescope on UT 2006 June 7--13.
The resulting spectra cover the wavelength range 3767--5227 \AA~with a dispersion of
0.235 \AA~pixel$^{-1}$ and resolution $R (\lambda/\Delta\lambda)=9580$.  Observations
were made for approximately four hours each night, and all the individual spectra from
each night were coadded to improve the S/N.  The coadded spectra were shifted to
a heliocentric frame and continuum normalized by fitting regions free of strong emission lines.

The observations were timed to correspond to mid-eclipse and greatest disk opening angle, 
as with the data from \citet{hil04}.  The spectra cover orbital phase $0.85\leq\phi\leq0.32$ 
and precessional phase $0.02\leq\psi\leq0.06$, according to the orbital ephemeris of \citet{gor98},
$$\mathrm{HJD}\ 2,450,023.62+13.08211E$$
and, for disk precession, the model ephemeris of \citet{gie02a},
$$\mathrm{HJD}~2,451,458.12+162.15E.$$
The orbital phase coverage overlaps with and extends that of \citet{hil04}. 

\section{Spectral Appearance and Radial Velocity Analysis}

The resulting spectra from seven consecutive nights around mid-eclipse show the expected
broad ``stationary'' emission lines.  The brightest emission lines of hydrogen and helium
all showed strong P Cygni absorption components every night except on the fourth and 
last nights of the run.  The H$\beta^-$ jet line was present in the spectra, but weak.

The new series of spectra appear to show considerably more emission contamination
in the range $\lambda\lambda$4500--4625 \AA ~than seen at the time of the 
investigation by \citet{hil04}, and consequently, it was not possible to 
measure unambiguously the same set of absorption lines they analyzed.  The only clearly
problem free regions larger than $\sim 25$ \AA~ in the observed spectral range are the 
intervals of $\lambda\lambda$4750--4830 \AA~and $\lambda\lambda$4950--4985 \AA .  These
regions show clear absorption features in the spectra.  An interesting case is that of
night seven, in which the only visible absorption features are interstellar in origin.
Their clear appearance in that spectrum allowed us 
to subtract the interstellar absorption from the spectra obtained on the first six nights.
This was especially helpful near 4762 and 4780 \AA~where interstellar 
lines overlapped apparent stellar absorption features.

\begin{figure}
\begin{center}
\epsscale{0.7}
\plotfiddle{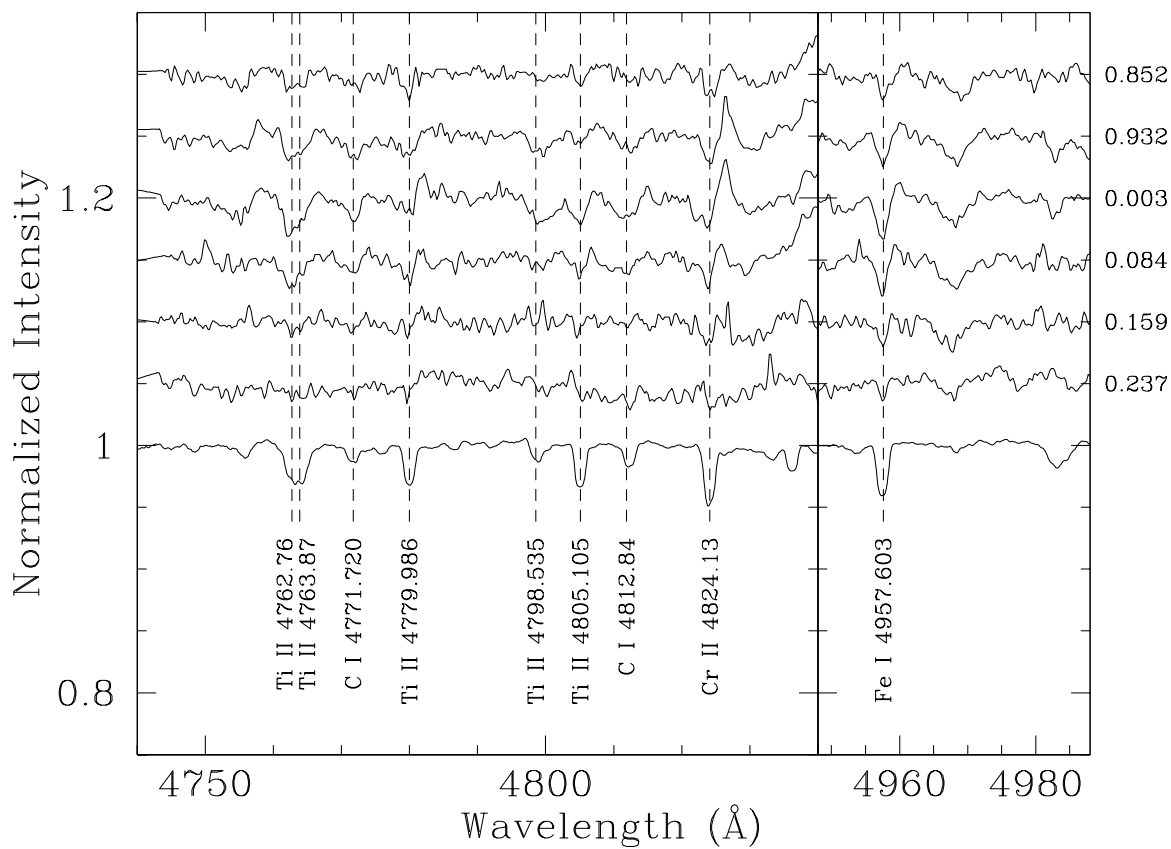}{100pt}{-90}{335}{450}{30}{0}
\end{center}
\caption[SS~433 mid-eclipse spectra] {SS~433 spectra obtained
with GMOS at Gemini North in 2006 June along with a spectrum
of the A4~Iab star HD~9233 ({\it bottom}) obtained with the 
MRS on HET in 2005 August.  The HD~9233 spectrum has
been smoothed and scaled to match the SS~433 spectra (see text).
Spectra have been offset for clarity.  The corresponding orbital phase
for each spectrum is shown on the right side of the plot.
\label{spec}}
\end{figure}

The resulting, interstellar-subtracted, and normalized spectra of SS~433 in the wavelength region
described above are shown in Figure \ref{spec} along with a spectrum of an A-supergiant, HD~9233 (A4~Iab).
The HD~9233 spectrum was obtained in 2005 August with the medium resolution
spectrograph (MRS) on the University of Texas Hobby-Eberly Telescope (HET).  
It has been smoothed to match the
resolution of the SS~433 spectra and intensity scaled to match the mid-eclipse
line depth as described below.  A comparison of the interstellar spectrum of SS~433 from 
night seven with that of HD~9233 shows that the latter has no appreciable interstellar absorption.
We confirm that the the lines of an A-type supergiant absorption spectrum are once
again present in the spectrum of SS~433.

Line identifications for major absorption lines are also shown in Figure \ref{spec}.
All the spectra in the figure are Doppler-shifted to the heliocentric rest frame.
The SS~433 spectra were shifted using the radial velocity values discussed below.
The HD~9233 spectrum was shifted using a measured 
heliocentric radial velocity of $-69.5\pm1.0$ km s$^{-1}$, significantly different
than the value of $-34\pm2$ km s$^{-1}$ given by \citet{hil04}.  It is likely that HD~9233
is a radial velocity variable.

\begin{figure}
\begin{center}
\epsscale{0.9}
\plotone{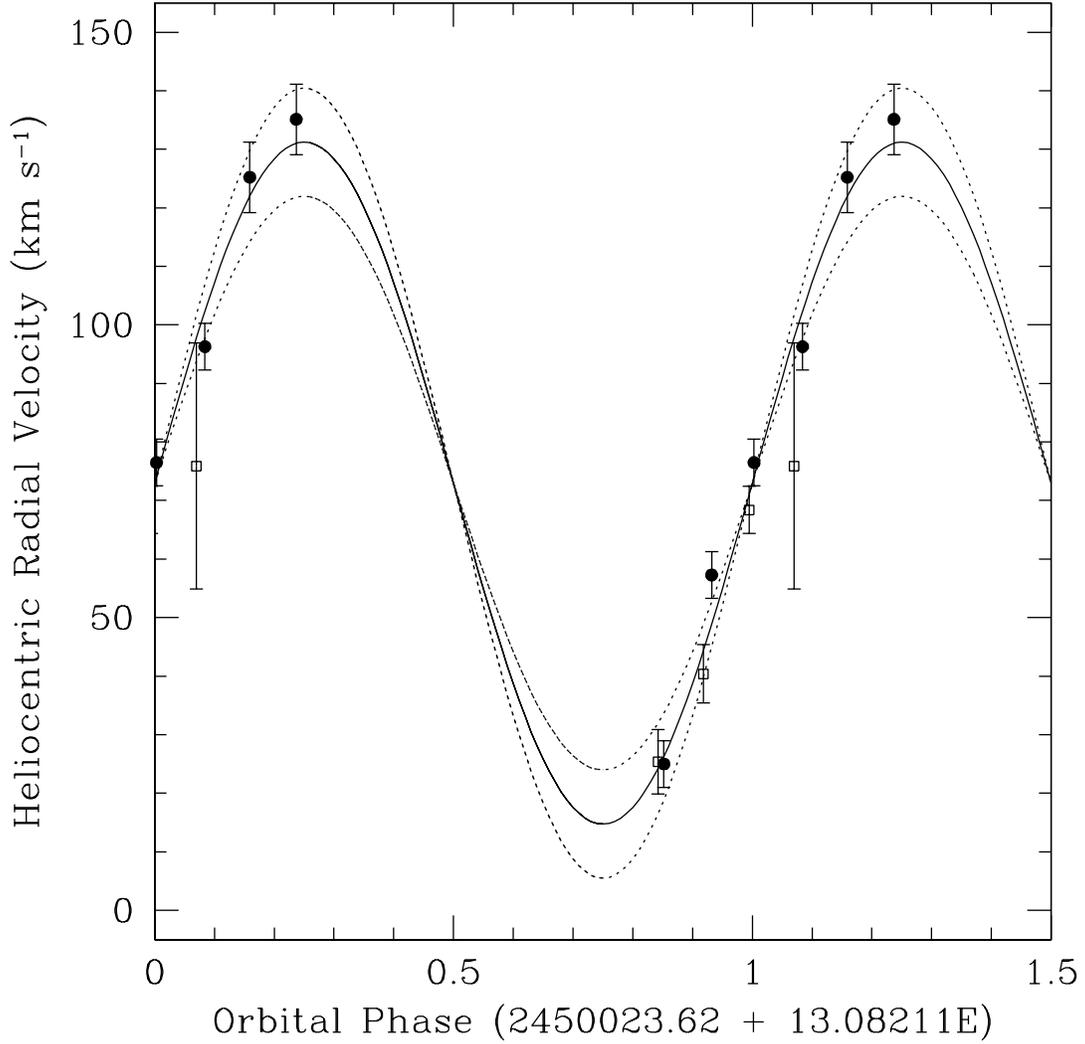}
\end{center}
\caption[SS~433 radial velocity curve] {The phase-folded
radial velocity curve for the absorption spectrum in SS~433.
Shown are the Gemini North data from 2006 June ({\it filled circles})
and the data from \citet{hil04} ({\it hollow squares}).  The dashed lines show the
$\pm 2\sigma$ semiamplitude solutions relative to the calculated best fit.
\label{ss433rv}}
\end{figure}

We measured radial velocities for the absorption lines in the 
SS~433 spectra using the same method applied by \citet{hil04}, i.e., 
by calculating cross-correlating functions (CCFs) of 
the absorption line regions in each SS~433 spectrum with those in 
the spectrum of HD~9233.  The spectrum of HD~9233 was first rescaled in intensity 
to match the absorption features of the mid-eclipse spectrum of SS~433.  The scaling factor
found in \citet{hil04} of $0.36\pm0.07$ was also found to be appropriate for this data.
We then added the radial velocity of HD~9233 to the relative 
velocities from cross-correlation to transform them to an absolute scale. 
The resulting heliocentric radial velocities for SS~433 are given in Table \ref{rvel} and
are plotted as a function of orbital phase in Figure \ref{ss433rv} (together with the  
radial velocity measurements from \citealt{hil04} for comparison).

\begin{deluxetable}{lcccc}
\tabletypesize{\footnotesize}
\tablewidth{0pc}
\tablecolumns{5}
\tablecaption{Absorption Line Radial Velocities\label{rvel}}
\tablehead{
\colhead {Date}        & 
\colhead {} & 
\colhead {} &
\colhead {$V_r$} & 
\colhead {$\sigma (V_r)$} \\
\colhead {(HJD - 2,450,000)}  & 
\colhead {$\psi$}      & 
\colhead {$\phi$}      &
\colhead {(km s$^{-1}$)} & 
\colhead {(km s$^{-1}$)} }
\startdata
3893.99\dotfill & 0.022 & 0.852  &   \phn25 &  4 \\
3895.04\dotfill & 0.029 & 0.932  &   \phn57 &  4 \\
3895.96\dotfill & 0.034 & 0.003  &   \phn77 &  4 \\
3897.02\dotfill & 0.041 & 0.084  &   \phn96 &  4 \\
3898.01\dotfill & 0.047 & 0.159  &      125 &  6 \\
3899.03\dotfill & 0.053 & 0.237  &      135 &  6 \\
3900.03\dotfill & 0.060 & 0.314  &  \nodata & \nodata \\
\enddata
\end{deluxetable}

The CCF fitting was adversely affected by the emission near the \ion{Cr}{2} $\lambda4824$ line
so that line was removed from the CCF fitting region.  Also, the fit on night six is clearly
dominated by the strongest line (\ion{Fe}{1} $\lambda4957$).  To determine if this line might be
consistently offset from the full sample CCF values, and thus offset the night six radial velocity
value, a CCF fit was run using just this line.  The resulting radial velocity and CCF amplitude
results for nights 1--6 were well within the individual error ranges of the full CCF fits
and there were no apparent systematic differences.

\begin{figure}
\begin{center}
\epsscale{0.9}
\plotone{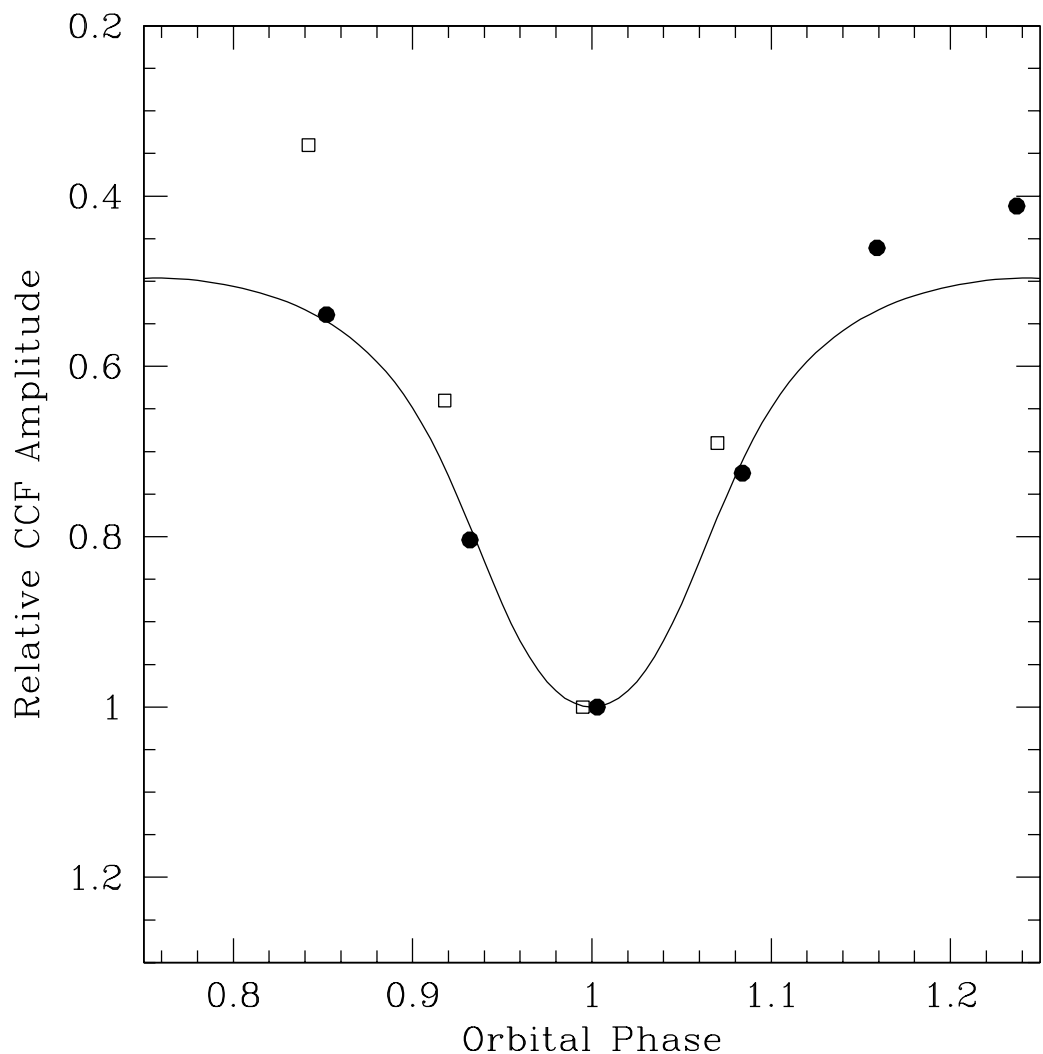}
\end{center}
\caption[SS~433 CCF amplitude curve] {The normalized 
CCF amplitude vs.\ orbital phase curve for the SS~433 absorption spectrum.
Shown are the results from the Gemini North data from 2006 June ({\it filled circles}),
the CCF data from \citet{hil04} ({\it hollow squares}), and the prediction ({\it solid line})
for a constant flux photospheric spectrum diluted by the variable flux from 
the accretion disk (according to the mean $B$-band light curve).
\label{ccf}}
\end{figure}

The absorption lines shown in Figure \ref{spec} appear to be strongest near mid-eclipse 
(phase 0.0) when the flux dilution from the disk light is minimized.   This variation 
in strength is represented in Figure \ref{ccf} by the relative CCF amplitudes for the
2006 June Gemini North data, along with those from \citet{hil04}.  The solid line 
in this figure shows the predicted variation in strength for a constant flux component 
that is diluted by the phase-variable flux contribution from the disk.   
This function was derived from the schematic $B$-band light curve of \citet{gor97,gor98} 
as described in \citet{hil04}.  Since the disk flux varies stochastically,
the actual eclipse light curve during the time 
of the 2006 June observations may have differed from this prediction.  
Nevertheless, it appears that the absorption lines did strengthen and 
weaken as expected during this particular eclipse. 

\section{Discussion}

\citet{hil04} discuss three criteria that must be met 
if the absorption line pattern is to be 
reliably associated with the supergiant:
(1) The radial velocity variation with orbital phase should 
appear with the ascending branch centered at orbital phase 
0.0 (i.e., at inferior conjunction at the time of 
the optical and X-ray mid-eclipse).  Furthermore, 
the systemic velocity should be approximately the same 
as that of the nebula ($+56$ to $+75$ km~s$^{-1}$; \citealt*{bou07,loc07}).
The new radial velocity data (Fig.~\ref{ss433rv}) agree with these 
predictions and with the earlier radial velocity measurements.
(2) The absorption line depths will usually be weak and modulated
in strength by the orbital curve and precessional phase.  
The observed spectrum is a composite of disk and stellar 
light, so that the stellar lines will always be diluted 
by the continuum flux from the disk.  Furthermore, the 
accretion disk is probably optically thick and vertically 
extended, so the supergiant will often be hidden when 
the star is behind the disk plane from our line of sight.
The new spectra show the same variation in absorption line
strength (Fig.~\ref{ccf}) as found previously by \citet{hil04},
consistent with our expectations for the varying supergiant to
disk flux ratio.
(3) The presence of a super-Eddington disk in SS 433 implies 
that the system is undergoing rapid mass transfer from 
an evolved and Roche-filling mass donor.  The fact that
the absorption line spectrum resembles that of an A-supergiant
with a projected rotational line broadening that matches 
expectations for a Roche-filling star in synchronous rotation 
\citep{hil04} is fully consistent with the mass transfer picture. 

All three criteria are met in the new observations, and the
remarkably consistent agreement between the new results and
those of \citet{hil04} strongly support the conclusion that 
the absorption pattern in the spectrum of SS~433 observed near 
mid-eclipse and precessional phase zero is that of the mass donor star.  
On the other hand, the behavior of an absorption spectral
component associated with a disk or disk wind will
be highly time variable, and we suspect that the kind of absorption 
features described by \citet{bar06} and \citet{cla07} do originate 
in a disk outflow as they propose. 

The radial velocity curve 
also provides us with the opportunity to revise the 
kinematical mass calculations for each component.
The best-fit radial velocity curve for the combined sets of data 
shown in Figure \ref{ss433rv} gives a semiamplitude for the 
mass donor star of $K_O = 58.2\pm3.1$ km s$^{-1}$ and a systemic
velocity of $\gamma = 73\pm2$ km s$^{-1}$.  Adopting a compact object
semiamplitude of $K_X = 168\pm18$ km s$^{-1}$ \citep[see][]{hil04} 
and system inclination of $78\fdg8$ \citep{mar89}
gives component masses of $M_O=12.3\pm3.3~M_\odot$ and $M_X=4.3\pm0.8~M_\odot$.
The larger revised mass for the compact object establishes it much more 
firmly as a black hole candidate.  The derived mass ratio $M_X/M_O=0.35$
is consistent with limits from the optical \citep{ant87} and X-ray 
\citep*{ant92,fil06} light curves.

We caution that our estimate of $K_O$ may differ from the actual geometric 
value if the inner hemisphere of the A-supergiant is significantly heated by
flux from the disk \citep{che05}.  The amount of irradiation present 
is very uncertain since the disk vertical extensions may be sufficient 
to block most of the high energy radiation from the central accretion 
zone from reaching the facing hemisphere of the supergiant.  
Furthermore, the effects of such heating on line formation are 
complicated.  For example, a hotter, brighter inner hemisphere will 
have a center of light that is shifted towards the companion (so that
the measured radial velocity shifts would be less than actual).  On the
other hand, if the temperature is high enough in the irradiated 
hemisphere to shift the plasma to higher ionization stages or to reduce 
the temperature gradient in the atmosphere, then the lines we observe 
would be weaker or absent in the surface elements of the inner 
hemisphere (shifting the measured radial velocities to larger than 
actual).  The clues from the available spectra suggest that irradiation 
probably plays a minor role since the decline in CCF strength following 
mid-eclipse follows approximately the expected curve for flux dilution
alone (Fig.~\ref{ccf}).
However, the absorption lines do weaken and disappear 
in the final spectrum obtained near orbital quadrature phase, 
and this behavior may be due to some contribution of irradiation or
the disappearance of the supergiant behind an 
optically thick accretion disk.  If irradiation of the inner hemisphere
were to change the ionization stage, however, then at quadrature the absorption
lines would be visible from only half of the projected stellar surface,
meaning the line strengths should be no greater than half that predicted by
the eclipse model ({\it solid line}) in Figure \ref{ccf}.

\section{Conclusions}

We have shown that the behavior of the A-supergiant absorption spectrum observed
in SS~433 during eclipse and near precessional phase zero is repeatable.
This suggests that the origin of the spectrum is a consistent, well-behaved source.
We suggest that the source is the mass donor star in the system rather than
the disk or a disk wind.  It is interesting
that SS~433 would show such similar spectra from both the mass donor star and
the disk/disk wind.  However, the appearance of such absorption might be 
anticipated if the disk outflow material is clumpy and has a temperature
in the line forming region comparable to that of the supergiant. 

The radial velocity curve resulting from observations of the mass donor star then allow
calculations of the kinematical masses of the two components.  We find masses of
$M_O=12.3\pm3.3~M_\odot$ and $M_X=4.3\pm0.8~M_\odot$ for the mass donor
and compact object.  It is possible that the inner hemisphere of the mass donor is
irradiated, which would change the mass of both components, potentially reducing
the compact object mass to below the neutron star limit.  The behavior of the
CCF amplitude suggests that a strong irradiation effect is not
present in the data presented here.  If this is the case, then the compact
object in SS~433 can be firmly identified as a black hole candidate.

\acknowledgments
We thank the staff of the Gemini North Observatory for their
help in obtaining the GMOS spectra described here. 
This material is based upon work supported by the 
National Science Foundation under Grant No.~AST-0606861 (TCH).
Institutional support has been provided from the GSU College
of Arts and Sciences and from the Research Program Enhancement
fund of the Board of Regents of the University System of Georgia,
administered through the GSU Office of the Vice President for Research (DRG).
We gratefully acknowledge all of this support.

{\it Facilities:} \facility{Gemini:Gillett (GMOS)}, \facility{HET (MRS)}


\end{document}